\documentclass[useAMS,usenatbib,usegraphicx]{mn2e}

\title[Predicting the incidence of planets and debris discs]
{Predicting the incidence of planets and debris discs as a function of 
stellar mass}
\author[J. S. Greaves]{J. S. Greaves$^1$\thanks{E-mail: jsg5 at 
st-andrews.ac.uk} \\
$^1$SUPA, School of Physics and Astronomy, University of St. Andrews, 
North Haugh, St. Andrews KY16 9SS, UK 
}

\begin{document}

\date{Accepted 2010. Received 2010; in original form 2010}

\pagerange{\pageref{firstpage}--\pageref{lastpage}} \pubyear{2010}

\maketitle

\label{firstpage}

\begin{abstract}

The mass of solids in a young circumstellar disc may be the key 
factor in its efficiency in building planetesimals and planetary 
cores, and dust observed around young T Tauri and Herbig Ae stars 
can be used as a proxy for this initial solid content. The dust-mass 
distributions are taken from recent millimetre-wavelength data and 
fitted using survival analysis to take into account upper limits, 
and threshold disc-masses for building planets and belts of comets 
are estimated. Amongst A-stars, 20~\% gas giant and 55~\% debris 
disc systems are predicted, in good agreement with observations. For 
M-stars, the predicted and observed planet-frequencies agree at 
$\sim 2-3$~\%, and this low incidence is explained by a lack of 
massive discs. However, debris is predicted around $\approx 14$~\% 
of M-stars, while only $\sim 2$~\% such systems have so far been 
found. This suggests that deeper searches such as with Herschel and 
SCUBA-2 may find a cold disc population previously missed around 
these low-luminosity stars. Also, an estimate of the efficiency of 
building millimetre-detected dust into planetary cores suggests that 
about a third of M-stars could host an Earth-mass planet -- but as 
the dust is spread over large disc areas, such planets may orbit far 
from the star.

\end{abstract}


\section{Introduction}

The incidence of extrasolar planets is known to be a function of 
both stellar mass and metallicity. If the present-day metallicity of 
the star reflects that of its circumstellar material at early times, 
then metal-rich discs would be expected to build planetary cores 
more readily by grain coagulation. The stellar mass may enter if 
more massive stars have higher-mass discs, as well as in factors 
such as shorter dynamical times for grains to collide. In the 
simplest approach, the mass of solids in a disc, i.e. the total mass 
M multiplied by the solid fraction Z, may be the relevant threshold 
for a successful outcome in forming planets or planetesimals 
\citep{wyatt07,greaves07}. These two quantities M,Z can be readily 
estimated using proxies. Millimetre-wavelength emission from dust is 
rather optically thin \citep{andrews07a} and the derived disc masses 
in present-day star formation regions are canonically multiplied by 
a factor $\approx 100$ to include the gas content inherited from the 
interstellar medium. These masses are proxies in the sense that 
material already converted into larger bodies (from boulders up to 
planets) has little emitting area, and so potentially only some of 
the orbiting material is represented. A good proxy for the 
refractory proportion of material in discs of non-solar metallicity 
is the logarithmic [Fe/H] of the host star.
 
\citet{greaves07} showed that such a simple empirical model can 
reproduce the properties of both planets and debris discs around 
Sun-like stars. The presence of debris indicates collisions between 
planetesimals, such as icy outer-system comets, treated in the model 
as a less successful mode of planet formation ending only in small 
bodies. Both the frequencies and metallicity-dependencies of these 
two outcomes were well matched. In particular, forming planets 
requires a high solid mass to build the core, and so high 
metal-fractions are advantageous, explaining the strong metallicity 
trend in the host stars \citep{wyatt07}. On the other hand, forming 
comet belts requires only modest amounts of material, and many discs 
meet this threshold regardless of metallicity, so it enters only 
weakly \citep{greaves07}. More complex approaches such as the 
simulations of \citet{johansen09} have confirmed that the outcomes 
of planetesimal formation should depend strongly on metallicity. 
\citet{raymond07} have used a model of disc-masses in solids to 
investigate the formation of terrestrial planets, finding in 
particular that habitable examples could be less likely for low-mass 
stars if they had insubstantial discs. A recent summary of our 
understanding of planetesimal growth is presented by 
\citet{chiang10}

\begin{figure}
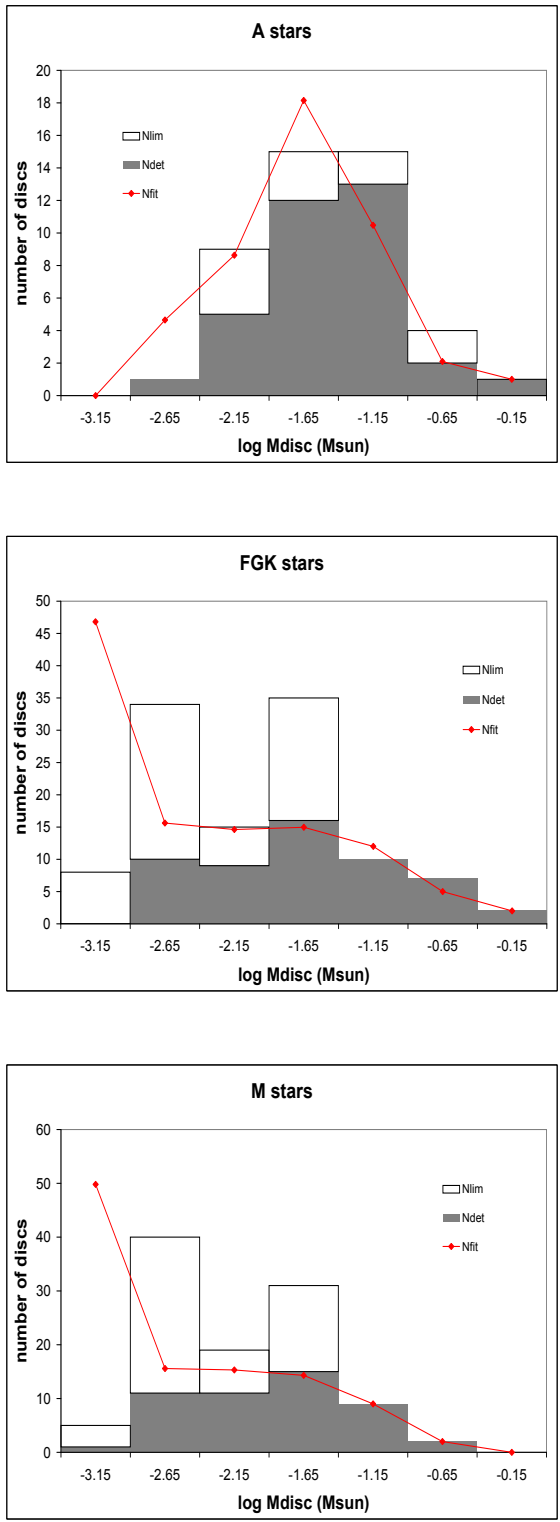

\label{fig1}
\includegraphics[height=85mm,width=70mm,angle=270]{KMestA.prn}
\includegraphics[height=85mm,width=70mm,angle=270]{KMestFGK.prn}
\includegraphics[height=85mm,width=70mm,angle=270]{KMestM.prn}
\caption{Distribution of disc masses (in gas plus dust) for A, FGK and 
M stars. Disc detections and upper limits are shown by filled and 
unfilled bars respectively; the red lines shows the fitted population 
from survival analysis (see text). X-axis labels are for the 
bin-centres. For reference, a solar mass is just over 1000 Jupiter 
masses; the lowest bin comprises discs of $<1.3$~M$_{Jupiter}$ or 
$<$4~M$_{Earth}$ in solids at solar metallicity. 
}
\end{figure}

Here our empirical model is expanded to test the planet and debris 
frequency expected for stars of higher and lower mass, namely A and M 
dwarfs. Their planetary systems are now being discovered by radial 
velocity surveys (of subgiant descendents in the case of A-stars), and 
their debris discs seen as far-infrared excesses e.g. by the Spitzer 
satellite \citep{su06,gautier07}. Here the M,Z distributions are 
assessed, and while the [Fe/H] ranges among the stars are all similar, 
the disc-mass distributions differ markedly, leading to strong changes 
in incidence with stellar mass.

\section{Data}

Metallicity data were first compared for different stellar types. 
Subgiants that are the descendents of A-stars of $\approx$~1.2-2.5 
solar masses can be identified by absolute magnitude M$_V$ of 0.5-3.5 
and B-V of 0.55-1.0 \citep{johnson06}, or similarly effective 
temperature 5000-6300~K \citep{flower96}. The SPOCS study of nearby 
stars \citep{valenti05} yields approximately 100 such subgiants with 
[Fe/H] measurements. A (log-)normal fit to these values indicates mean 
and standard deviation $\mu,\sigma$ of +0.02, 0.23. This is very 
similar to values for nearby Sun-like dwarfs, with $\mu,\sigma$ of 
-0.01,0.29 in the data of \citet{valenti05}. For M-stars, the 
derivation of metallicities based on photometry has recently be 
re-computed by \citet{johnson09}. Using data they present for M-stars 
within 20 pc, and their relation for [Fe/H] derived from K-magnitude 
height above the main-sequence, $\mu,\sigma$ are found to be 
-0.06,0.24. These distributions\footnote{\citet{schlaufman10} argue 
that the mean [Fe/H] for FGK and M stars are about 0.1 dex lower than 
found here; such small shifts have negligible effect on the 
M$_{solids}$ distribution.} are so similar that they can not 
significantly affect planet-building potential around different classes 
of star.

\begin{table*}
 \caption{Predicted and observed percentages of giant planets and 
debris discs. The planet data are from the uniform-detectability 
sample of \citet{johnson10} and debris frequencies are from 
Spitzer 70~$\umu$m surveys \citep{gautier07,trilling08,beichman06,su06},  
limited to data with signal-to-noise$> 3$. The debris frequency observed 
for FGK stars is used to make the predictions for other spectral types. 
The upper limit for M-stars is from $<$1/15 objects detected by Spitzer 
at 70~$\umu$m in $\geq 3\sigma$ data \citep{gautier07}, while the bracketed 
value is from 1/50 detections in the millimetre survey of \citet{lestrade09}. 
Poisson errors are based on the number of planet and debris systems 
detected, and the numbers of discs falling within the input mass-bounds 
(see text). The parameters of Equation~1 are listed in the four columns 
after the spectral types. The A-star mass marked with a $^*$-symbol is 
actually the second most-massive value from \citet{natta04}, as 
another disc with large distance and uncertain mass was removed. 
The sample of \citet{alonso09} gives a comparable highest mass of 
225 Jupiter masses among HAe stars within about 500~pc.
}
 \label{tab:list}
 \begin{tabular}{@{}ccccccccc}
  \hline
spectral type & max $M_{disc}$& min. [Fe/H]   & min. $M_{solids}$ & min. $M_{disc}$ & \multicolumn{2}{c}{f(planet) \%} & \multicolumn{2}{c}{f(debris) \%} \\
              &	(M$_{Jup}$)   &	(planet)      &	(M$_{\oplus}$, for planet) & (M$_{Jup}$, at [Fe/H]=0)   & predict& observe & predict & observe \\
  \hline
FGK	& 600 	& -0.7	& 400	& 120 	& 8 $\pm$ 3	& 8.5 $\pm$ 1.0 & --	& 	19 $\pm$ 3 \\
A	& 150$^*$& -0.36& 200 	& 60	& 20 $\pm$ 7	& 20 $\pm$ 3	& 55 $\pm$ 11	& 70 $\pm$ 11 \\
M	& 270	& -0.12	& 650 	& 200	& 1.9 $\pm$ 1.3	& 3.3 $\pm$ 1.5	& 14 $\pm$ 4	& $< 7$	
($\sim 2$)\\
  \hline
 \end{tabular}
\end{table*}

The disc mass distributions were then constructed and found to be quite 
different for A versus FGK and M stars (Figure~1). The masses for the 
A-stars are from \citet{natta04}, in a millimetre-based compilation for 
young Herbig Ae objects\footnote{The data presented by \citet{alonso09} 
give a similar mass-distribution but with fewer HAe stars.}. Their plot 
of M$_{disc}$/M$_{star}$ versus M$_{star}$ was used to derive 
M$_{disc}$ for M$_{star}$ of 1.2-2.5 solar masses. Since 11/45 of the 
disc-mass measurements were upper limits, a survival analysis was used 
to find the underlying distribution, implemented with the {\sc asurv} 
package \citep{lavalley92}. The resulting disc-mass distribution 
appears rather log-normal in form, and is centred at a mean of 25 
Jupiter masses of gas and dust. For comparison, this is just above the 
Minimum Mass Solar Nebula of $\approx$20 Jupiter masses \citep{davis05} 
needed to supply the cores of all the Sun's planets. For FGK and M 
stars\footnote{Pre-main sequence M-stars have rather vertical Hayashi 
tracks with modest change in effective temperature, so young and old 
objects of M-type will have similar stellar masses. Other pre-main 
sequence stars evolve towards hotter spectral types, but since most of 
the observed T Tauri objects have K-classifications, they will still be 
within roughly `Sun-like' classes at later times.}, the 
millimetre-surveys of Tau/Aur and Oph by \citet{andrews05,andrews07b} 
were used, comprising similar-aged stars to the HAe's of 
\citet{natta04}. Since these authors used a conversion of flux to dust 
mass yielding values a factor of three lower than in \citet{natta04}, 
the masses have been shifted upwards by one bin ($\delta$log-M = 0.5) 
in Figure~1. The difference arises from using an opacity at 1~mm 
wavelength of 0.03~cm$^2$/g rather than the 0.01~cm$^2$/g found in the 
detailed grain model of \citet{natta04}, which also agrees with recent 
literature values \citep{draine06}. Survival analysis was again used to 
construct the base disc-mass populations for FGK and M stars, but more 
of the points are now upper limits, as the discs are less massive than 
for the HAe stars . There were 57 limits out of 111 measurements for 
the FGK dwarfs, and 57/106 for the M-stars. The means are 4 and 3 
Jupiter masses respectively. There is also a large tail of 
insubstantial discs in both cases, which combined with the lower means, 
suggests that these discs will have much less planet-forming capability 
than those of A-stars. The low-mass tails include many Class~III T 
Tauri stars, which may lose their discs early on \citep{luhman10}. All 
this disc-mass distributions are very broad, whereas the metallicity 
distributions would cover only about one bin on the scales of Figure~1.

\section{Results and predictions}

\subsection{Planets}

Since the metallicity distributions are all narrow and similar, the 
differences in M$_{disc}$ should be the dominant factor, and the 
relevant thresholds for planet formation are now estimated. 
\citet{greaves07} argued that if a minimum solid mass is needed to 
grow a planetary core, then there will be observed examples of giant 
planets arising from a range of discs with this same solid-mass, but 
ranging between massive/low-metal and lower-mass/high-metal systems. 
The lowest-metal system observed with a planet arises when the 
solids criterion was met only because the disc was at the extreme 
high-end of total mass. The solid mass can then be expressed using 
the proxy data as
\begin{equation}
M^{min}_{solids} = M^{max}_{disc} 10^{\rm [Fe/H]_{min}}/100. 
\end{equation}

The resulting parameters for all the spectral types are listed in 
Table~1. The input data on the disc masses are from the references 
in the previous section, and the metallicities of planet-hosts are 
from http://exoplanet.eu/. The subsequent columns list the disc-mass 
thresholds needed to form planets, and the predicted and observed 
frequencies. For FGK stars, the predicted incidence of around 8~\% 
is in excellent agreement with the 8.5~\% found in the systematic 
planet sample of \citet{johnson10}. For A-stars the planet 
agreement is also very good, with around 20~\% both predicted and 
observed. For M-stars, the incidence is 2-3~\%, and the model 
correctly predicts a low value because few of
the discs are massive (Figure~1).

The solid-mass thresholds needed for forming giant planets are here 
found to be 200, 400 and 650~M$_{\oplus}$ for A, FGK and M spectral 
types respectively (Table~1). This trend qualitatively matches the 
expectation that grains will coagulate more readily in discs around 
more massive stars, as the dynamical (orbital) times are shorter.

Subsets of different types of planetary systems are not considered 
in detail here. Rocky `super-Earth' planets form a small part of the 
FGK-hosted population (e.g. $\la10$~M$_{\oplus}$ bodies contribute 
around 0.5~\% incidence), and the requirements for building such 
bodies are considered briefly below. Also, there are likely to be 
systems with more gas giants orbiting further out -- for example, 
\citet{gould10} estimate $\sim1/6$ for the frequency of two-giant 
systems, from microlensing results -- but these are not included 
here. The analysis is however self-consistent in using the 
lowest-metal radial-velocity detection to predict the planet 
frequency in this type of survey.

\subsection{Debris discs}

For the less-successful outcome of building planetesimals, the 
threshold can not be straightforwardly derived. \citet{greaves07} 
used the metallicity range of FGK debris-hosts to estimate the 
mass-threshold for building comet belts, but there is little similar 
data for A and M-stars. Hence the simplest procedure is to infer the 
threshold from the data. In deep Spitzer surveys, approximately 
19~\% of FGK stars have debris discs detected at 70~$\umu$m 
\citep{greaves10}, associated with belts of colliding planetesimals. 
Associating this outcome with discs of lower-mass than those that 
build planets, this next 19~\% of the M$_{disc}$ distribution 
corresponds to $>$50 Earth masses in solids or $>$15 Jupiter masses 
in total at solar metallicity. This is eight times lower than the 
value needed for building planets. Making the minimal assumption 
that the efficiency of the planetesimal-building process is similar 
in all systems, a similarly reduced value can be applied for the 
other stellar masses. Discs of at least 25 and 80 Earth masses in 
solids would then be needed for A and M stars to have debris, 
respectively.

The model results (Table~1) show a predicted debris incidence 
for A-stars that is slightly low, but only at the -1$\sigma$ bound 
(error on observed$-$predicted of 15~\% in quadrature). For M-stars, 
there is however an interesting discrepancy between the debris 
prediction of around 14~\% and the observed incidence of as little 
as 2~\% (Table~1). A few other M-host debris discs have been 
discovered, but all around stars much younger than typical 
main-sequence lifetimes \citep{liu04,forbrich08, plavchan09}. It is 
possible that these M-discs evolve down to lower dustiness than seen 
among A or FGK stars (Figure~2), but given that FGK and M discs are 
similar at the T Tauri phase, an alternative explanation is 
favoured. As M-dwarfs are of low luminosity, dust particles in 
thermal equilibrium with the star will be cool, compared to a grain 
on the same orbit around AFGK stars. In this case, the discs will 
have faint emission shifted to long wavelengths, and this introduces 
a survey bias \citep[e.g.]{matthews07}. There may be negligible 
excess above the M-star photospheres at 70~$\umu$m, and exploratory 
millimetre surveys may not yet have reached deep enough to detect 
the discs.

There are some systems known with both gas giants and debris discs, 
but the statistics suggest the two phenomena are not correlated 
\citep{bryden09}; here in fact it is assumed that they are separate 
outcomes. In the uncorrelated case for FGK stars, a proportion of 
0.08 (planet-frequency) $\times$ 0.19 (debris frequency) or 1.5~\% 
of stars would have planets and dust, so this should be a negligible 
population. A small number of planets around A-hosts of debris discs 
has also been found in direct imaging surveys. These cases are not 
considered as a separate class here. The stability of debris discs 
over several Gyr, in terms of size and mass and largest bodies 
present, has been studied in detail by \citet{heng10}.

\begin{figure}
\label{fig2}
\includegraphics[width=88mm,angle=0]{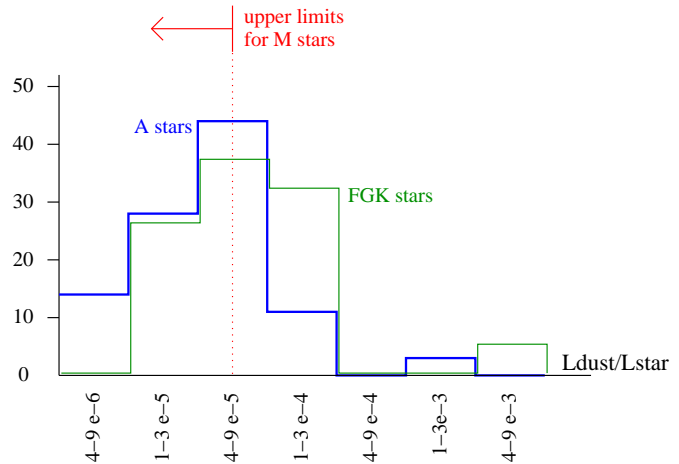}
\caption{Distributions of percentage of stars versus fractional dust 
luminosities for A and FGK systems with debris detected at 
24,70~$\umu$m \citep{su06,trilling08, beichman06}. The typical upper 
limit for the M-stars observed by \citet{gautier07} is shown at the 
top, and applies for dust at temperatures of $\approx$~40-100~K. 
}
\end{figure}

\subsection{Planet-building efficiency}

The planetary systems of M-dwarfs are of particular interest because 
these stars are so abundant in the Galaxy. The large solid-threshold 
inferred for M-stars to form planets (Table~1) suggests the core 
accretion process may be inefficient. A rough efficiency estimate 
can be made, assuming the solid content in the giant planets in the 
M-star sample of \citet{johnson10} is of order 20~M$_{Earth}$. For 
example, these planets are of $\ga$~M$_{Saturn}$ and this is about 
Saturn's solid content \citep{saumon04}. A lower-mass `super-Earth' 
population is also being discovered of which the most massive 
observed in transit has a 20~M$_{Earth}$ core plus $\approx$10-15~\% 
atmospheric mass \citep{bean08}. If 20 Earth-masses is a rough 
boundary above which planets got on to become gas giants, then we 
could equate this outcome at $\sim3.3$~\% frequency with the same 
fraction of top-end discs, which have $\ga 300$~M$_{\oplus}$ of 
dust. In this case, the efficiency of incorporating disc solids (or 
at least the millimetre-detected component) is $\sim 20/300$ or 
7~\%.

One consequence is that to build an Earth-mass planet, 
15~M$_{\oplus}$ of dust in the disc would be needed. This occurs for 
around 1/3 of M-stars (Figure~1), so they are not all expected to 
form Earth-analogues. Given that about 3/4 of main-sequence stars 
are of M-type, at least a quarter of nearby stars could host an 
Earth-mass planet under this simple scaling approach. However, the 
disc material spreads far from the star \citep{andrews07b}, so these 
planets may only be able to form in large orbits, particularly as 
the Hill radius for accretion increases with radial distance. 
Microlensing is particularly well suited to discovery of M-dwarf 
systems hosting such low-mass planets in orbits out to around 10~AU 
\citep{gould10}.

It is beyond the scope of this study to assess how many planets 
might migrate inwards, to the the habitable zone around 0.1~AU where 
temperatures are suitable for liquid water on the surface. 
\citet{raymond07} have simulated the formation of terrestrial 
planets, and find that habitable planets of $\ga0.3$~M$_{\oplus}$ 
(the boundary for plate tectonics and atmospheric retention) are 
much less likely for M-dwarfs than G/K stars. Figure~1 also suggests 
that discs substantial enough to form an Earth-mass planet are much 
more common around A stars, although these discs may also form 
giants, with implications for the supply of material and stability 
for terrestrial bodies \citep[e.g.]{fogg09}.

\section{Conclusions}

The disc-mass distributions for young A and M stars have been 
constructed and compared to those for Sun-like stars. Inferring the 
mass-thresholds for forming extrasolar planets and debris discs, the 
higher incidence of both phenomena among A-stars or their descendents 
is well reproduced, as is the low incidence of giant planets around 
M-stars, which rarely have substantial discs. However, only about 2~\% 
of nearby M-stars have debris discovered so far, and several times 
more are expected. There may be a population of cool debris discs 
around red dwarfs largely missed in surveys so far. The new SCUBA-2 
camera \citep{holland06} observing at 450 and 850~$\umu$m will be 
ideally matched to cold dust, as is the highly sensitive Herschel 
satellite \citep{pilbratt10}, now observing out to 500~$\umu$m. A 
further prediction of the empirical calculations is that about a third 
of M-stars have the capability to form a 1 Earth-mass planet -- 
however, such planets are likely to form far from the star, as the disc 
material is very spread-out.

\section*{Acknowledgments} 

I thank Sean Raymond for very useful discussions, an anonymous 
referee for helpful comments that improved the clarity of the paper, 
and STFC for a fellowship in support of this work.

\bsp

\label{lastpage}

\end{document}